\documentclass[%
	reprint,
	superscriptaddress,
	amsmath,amssymb,amsfonts,
	aps,
	pra,
]{revtex4-2}

\usepackage[utf8]{inputenc}
\usepackage[T1]{fontenc}
\usepackage{lmodern}
\usepackage[english]{babel}
\usepackage[autostyle=true]{csquotes}

\usepackage[dvipsnames]{xcolor}
\usepackage{graphicx}
\graphicspath{{figures/}}

\usepackage{booktabs}
\usepackage{tabularx}
\newcolumntype{L}{>{\raggedright\arraybackslash}X}
\newcolumntype{Y}{>{\centering\arraybackslash}X}
\newcolumntype{R}{>{\raggedleft\arraybackslash}X}

\usepackage{hyperref}
\hypersetup{%
}

\usepackage{tikz}
\usepackage{tikz-cd}
\usetikzlibrary{quantikz2}
\usetikzlibrary{%
  matrix,%
  calc,%
  arrows%
}

\usepackage{bm}
\usepackage{mathtools}
\usepackage{braket}

\usepackage{xspace}
\usepackage{mleftright}
\mleftright

\usepackage[%
	caption=false,
	subrefformat=simple,
	labelformat=simple,
]{subfig}

\usepackage[
	capitalise,
]{cleveref}

\usepackage[%
    separate-uncertainty=false,
    range-phrase=-,
]{siunitx}

\DeclareSIUnit\bohr{\text {\ensuremath {a}}_{0}}

\renewcommand{\vec}{\bm}

\newcommand*{\diff}{\mathop{}\!\mathrm{d}}
\newcommand*{\etothepowerof}[1]{\mathrm{e}^{#1}}
\newcommand*{\etothe}[1]{\etothepowerof{#1}}

\DeclarePairedDelimiter{\halfopeninterval}{[}{)}
\DeclarePairedDelimiter{\closedinterval}{[}{]}

\DeclarePairedDelimiter{\abs}{|}{|}

\DeclarePairedDelimiter{\ketx}{\lvert}{\rangle}
\newcommand*{\ensembleaverage}[1]{\overline{#1}}
\DeclareMathOperator*{\round}{round}

\DeclarePairedDelimiterX{\braketx}[3]{\langle}{\rangle}{#1\,\delimsize\vert\,\mathopen{}#2\,\delimsize\vert\,\mathopen{}#3}

\DeclarePairedDelimiterX{\setx}[1]\{\}{%

#1
}

\newcommand*{\algorithm}[1]{\ensuremath{\mathsf{#1}}\xspace}
\newcommand*{\complexityclass}[1]{\ensuremath{\mathsf{#1}}\xspace}
\newcommand*{\Hcost}{\ensuremath{\hat{H}_{\mathrm{cost}}}\xspace}
\newcommand*{\Hdrive}{\ensuremath{\hat{H}_{\mathrm{drive}}}\xspace}
\newcommand*{\sweepduration}{\ensuremath{t_{\mathrm{sweep}}}\xspace}
\newcommand*{\annealingtime}{\ensuremath{T}\xspace}
\newcommand*{\minimumgap}{\ensuremath{\Delta E}\xspace}

\definecolor{mypurple}{rgb}{0.49,0.18,0.56}
\definecolor{mygold}{rgb}{0.93,0.49,0.13}
\definecolor{mygreen}{rgb}{0,0.5,0}
\definecolor{myblue}{rgb}{0,0,0.75}
\definecolor{mymagenta}{cmyk}{0,1,0,0.12}
\definecolor{mygray}{rgb}{0.5,0.5,0.5}

\usepackage[normalem]{ulem}

\newif\ifcomments
\commentstrue 

\begin{document}
\title{Boosting quantum annealing performance through direct polynomial unconstrained binary optimization}

\author{Sebastian Nagies}
\email{sebastian.nagies@unitn.it}
\thanks{Corresponding author}
\affiliation{Pitaevskii BEC Center and Department of Physics, University of Trento, Via Sommarive 14,
38123 Trento, Italy}
\affiliation{INFN-TIFPA, Trento Institute for Fundamental Physics and Applications, Trento, Italy}

\author{Kevin T. Geier}
\affiliation{Pitaevskii BEC Center and Department of Physics, University of Trento, Via Sommarive 14, 38123 Trento, Italy}
\affiliation{INFN-TIFPA, Trento Institute for Fundamental Physics and Applications, Trento, Italy}
\affiliation{Quantum Research Center, Technology Innovation Institute, P.O. Box 9639, Abu Dhabi, United Arab Emirates}

\author{Javed Akram}
\affiliation{eleQtron GmbH, Heeserstr.\ 5, 57072 Siegen, Germany}

\author{Dimitrios Bantounas}
\affiliation{eleQtron GmbH, Heeserstr.\ 5, 57072 Siegen, Germany}

\author{Michael Johanning}
\affiliation{eleQtron GmbH, Heeserstr.\ 5, 57072 Siegen, Germany}

\author{Philipp Hauke}
\email{philipp.hauke@unitn.it}
\thanks{Corresponding author}
\affiliation{Pitaevskii BEC Center and Department of Physics, University of Trento, Via Sommarive 14, 38123 Trento, Italy}
\affiliation{INFN-TIFPA, Trento Institute for Fundamental Physics and Applications, Trento, Italy}

\begin{abstract}
Quantum annealing aims at solving optimization problems of practical relevance using quantum-computing hardware. Problems of interest are typically formulated in terms of quadratic unconstrained binary optimization (QUBO) Hamiltonians. 
However, many optimization problems are much more naturally formulated in terms of polynomial unconstrained binary optimization (PUBO) functions of higher order. 
As we show with various problem examples, leveraging the PUBO formulation can bring considerable savings in terms of required number of qubits. 
Moreover, in numerical benchmarks for the paradigmatic 3-SAT problem, we find scenarios where the scaling of the minimum energy gap during the optimization sweep differs significantly, suggesting the possibility of an exponentially faster annealing time when using the PUBO as compared to the QUBO formulation.
This advantage persists
even when considering the overhead caused by the higher-order interactions necessary for PUBO cost Hamiltonians.
As an interesting side effect, the analysis on minimum energy gaps of different 3-SAT instance generators reveals different degrees of hardness, which will be of interest also for classical benchmark calculations. 
Our findings show a promising path to improving the resource efficiency and sweeping speed of quantum annealing protocols on both analog and digital platforms, which are important prerequisites when aiming at solving larger optimization problems with relevance to industry. 
\end{abstract}

\maketitle

\section{Introduction}\label{sec:intro}

The continuous improvements of quantum computing devices have positioned quantum annealing as a promising candidate for tackling combinatorial optimization problems of practical relevance~\cite{Santoro2002, Martonak2004,Hauke2020, Rajak2022, Yarkoni2022}. Experiments on commercially available quantum annealers have recently even claimed quantum advantage~\cite{King2024}. 
Since most of the existing quantum computers consist of discrete two-level systems coupled by two-qubit gates, the most common approach for solving combinatorial optimization problems on quantum devices is via quadratic unconstrained binary optimization (QUBO), where the cost function is a polynomial of degree two.
However, many problems are more naturally expressed as polynomials of higher degrees, i.e., in terms of polynomial unconstrained binary optimization (PUBO), sometimes also referred to as higher-order binary optimization (HOBO). Application-relevant examples of PUBO problems range from vehicle routing~\cite{Weinberg2023} to polymer sampling~\cite{Micheletti2021,Slongo2023} and lattice protein folding~\cite{Outeiral2021,Chermoshentsev2021}. 
To implement these on quantum hardware, the PUBO problems are usually reduced to QUBO problems~\cite{Hauke2020,Schmidbauer2024}, and efforts have been made to render the reduction of PUBO to QUBO instances as efficient as possible~\cite{Biamonte2008,Babbush2013}. 
Nevertheless, the QUBO reduction comes at the price of introducing additional ancillary variables, along with constraints that enforce their consistency, both of which is best avoided considering the stringent space and interaction-scale limitations of current NISQ devices. 
Then again, considering the overhead incurred by the associated multi-qubit interactions, i.e., slower interaction rates on analog platforms or increased circuit depth on digital architectures, it remains an open question whether a direct PUBO implementation could be advantageous in quantum optimization~\cite{Stein2023}.  

Here, we answer this question in the affirmative, by highlighting the potential for a gain in efficiency through a direct implementation of PUBO problems. 
Paradigmatic problems that are naturally formulated in terms of PUBO range from the Boolean satisfiability problem (SAT), over the hypergraph coloring problem, to the $p$-spin model. 
Computing the number of qubits necessary to implement such problems as QUBO instances, we find a potential resource saving of up to an order of magnitude in the number of qubits for some problems when formulating them in terms of PUBO problems instead. 
Additionally, we compute the minimum energy gap across a range of problems by employing exact diagonalization on small-scale systems, specifically 3-SAT instances of varying complexity generated using different methods~\cite{Hsieh2021}. 
We find the direct formulations as PUBO instances to have larger gaps than the equivalent QUBO reductions,
suggesting an improvement in computing time.
While for hard instances the gap closing and thus the time-to-solution nonetheless scale exponentially with the problem size, direct PUBO formulations can exhibit smaller exponents for some classes of 3-SAT problems, which translates to an exponential speedup over the corresponding QUBO formulation.
This behavior is agnostic to the underlying quantum computing architecture and holds for analog as well as digital platforms alike.

For gate-based quantum computers, we discuss the possibility to synthesize the necessary 3-body spin interactions  with typical gate sets employing single-qubit gates combined with CNOT or $R_{zz}$ gates. The required overhead of only four two-qubit gates per three-qubit gate is more than compensated by the saving of ancillary qubits as well as the increased minimum energy gap.
This shows that it can be of large practical advantage to directly implement PUBO formulations in quantum annealing protocols, thus increasing the scalability of these machines.

On top of that, our studies on the scaling of the minimum energy gap also shed light on the hardness of instances created by various 3-SAT benchmark generators~\cite{Hsieh2021}.
Among the tested generators, we find completely randomly generated instances
to be on average easier to solve (larger minimum energy gap) than the structured instances with fewer logical clauses
constructed by one of the benchmark generators. 
In contrast, for yet another benchmark generator,
the formulation as a PUBO shows that it generates only trivially solvable instances, reflected in a large energy gap.
Thus, our analysis of solving 3-SAT on quantum computers can also inform benchmark calculations for combinatorial optimization problems on classical computers, where recent studies have demonstrated significant performance gains through quantum-inspired metaheuristic solvers for problems as large as $\num{100000}$ variables~\cite{Punnen2022,Glover2022,Du2025}.

This paper is organized as follows.
\Cref{sec:annealing} contains a brief recap on solving combinatorial optimization problems with quantum annealing and its main limitation due to the closing of the minimum energy gap.
The behavior of the minimum gap will be our main figure of merit for assessing potential performance gains of PUBO over QUBO.

 In \cref{sec:pubo}, we discuss the reduced spatial resource requirements (i.e., number of qubits) of optimization problems that are naturally formulated as PUBO, as compared to equivalent QUBO reductions.
\Cref{sec:numerics} puts forward our numerical results on the behavior of the minimum energy gap with the example of the paradigmatic 3-SAT problem, illustrating the potential for PUBO speedups due to larger gaps.
In \cref{sec:3body}, we discuss the synthesis of three-body interactions using single- and two-qubit gates, and compare the resulting overhead with the expected performance gain. 
Finally, \cref{sec:conclusion} contains our conclusions on the potential advantages of PUBO over QUBO in terms of spatial and temporal resources.

\section{Quantum annealing for polynomial unconstrained binary optimization\label{sec:annealing}}

In this section, we introduce the PUBO and QUBO formulations of combinatorial optimization problems, how these can be solved using quantum annealing, and how we assess the potential advantage of PUBO over QUBO in this study.

\subsection{Polynomial unconstrained binary optimization}

Quantum annealing aims at finding the optimal solution to a combinatorial problem~\cite{Hauke2020, Rajak2022, Yarkoni2022}.

Such optimization problems can be posed as finding the minimum of a cost function $f: \set{0, 1}^N \to \mathbb{R}$, mapping a Boolean vector (or bit string) $\vec{x}=(x_1, \dots, x_N)$ to a real number.
The problem falls into the realm of PUBO if the cost function can be written as a (multivariate) polynomial,
\begin{equation}
    f_{\mathrm{PUBO}}(\vec{x}) = \ \sum_{\mathclap{i_1 + \cdots + i_N \le P}} \ a_{i_1 \dots i_N} x_1^{i_1} \dots x_N^{i_N} \,,
\end{equation}
where $a_{i_1 \dots i_N}$ are coefficients, $P$ is the total degree of the polynomial, and the indices vary over non-negative integers.
This comprises QUBO as the special case where the cost function is quadratic ($P = 2$),
\begin{equation}
    f_{\mathrm{QUBO}}(\vec{x}) = \sum_{i, j = 1}^N Q_{ij} x_i x_j + \sum_{i = 1}^N c_i x_i \,,
\end{equation}
where $Q$ is a symmetric matrix, $\vec{c} = (c_1, \dots, c_N)$ is a vector of local biases, and we generally discard the irrelevant constant term.

Equivalently, the problem can be formulated in terms of classical discrete Ising spin variables $\vec{s} = (s_1, \dots, s_N)$, taking values in $\set{+1/2, -1/2}^N$.
Spin and logical variables are related to each other via the one-to-one mapping $\vec{s} = 1/2 - \vec{x}$.
Note that this mapping does not change the degree of the optimization problem (i.e., PUBO stays PUBO).
Solving the minimization problem then corresponds to finding the ground state configuration of the associated spin system which minimizes the Hamiltonian function $H_{\mathrm{cost}}(\vec{s}) = f(1/2 - \vec{s})$.

\subsection{The quantum annealing paradigm}\label{subsec:annealing_paradigm}

In quantum annealing, the classical spins are promoted to quantum spins (or qubits) by identifying them with spin-$1/2$ operators, $s_i \to \hat{s}_i^z$ (note the circumflex denoting operators).
The cost function (or Hamiltonian function) now corresponds to an operator, the cost Hamiltonian~$\Hcost = f(1/2 - \vec{\hat{s}})$, which is diagonal in the computational basis.
For the purposes of this work, we will mainly be interested in PUBO with cubic cost functions, where the associated cost Hamiltonians are of the form
\begin{equation}
\label{eq:3SAT_PUBO}
    \Hcost = - \sum_{ijk} J_{ijk}^{(3)} \hat{s}^{z}_{i} \hat{s}^{z}_{j} \hat{s}^{z}_{k} - \sum_{ij} J_{ij}^{(2)} \hat{s}^{z}_{i} \hat{s}^{z}_{j} - \sum_{i} h_{i}^z \hat{s}^{z}_{i} \,.
\end{equation}
Here, $J^{(3)}$ and $J^{(2)}$ represent three-body and two-body spin interactions, respectively, while the vector $\vec{h}^z = (h_1^z, \dots, h_N^z)$ describes a (typically inhomogeneous) longitudinal field.

The idea of quantum annealing is to evolve the system from the ground state of a suitably chosen driving Hamiltonian~$\Hdrive$ to the desired ground state of the cost Hamiltonian~$\Hcost$ by adiabatically changing Hamiltonian parameters.
In the simplest case of a linear sweep from $\Hdrive$ to $\Hcost$, the instantaneous Hamiltonian reads
\begin{equation}
\label{eq:annealingprotocol}
    \hat{H}(s) = (1 - s) \Hdrive + s \Hcost \,,
\end{equation}
where $s = t / \sweepduration \in \closedinterval{0, 1}$ is the time in units of the sweep duration~$\sweepduration$.
The driving Hamiltonian should not commute with the cost Hamiltonian and its ground state should be easy to prepare.
In practice, the driving Hamiltonian is therefore often chosen as $\Hdrive = - h^x \sum_i \hat{s}_i^x$, where $\hat{s}_i^x$ is the spin-$1/2$ operator on the $i$-th spin in $x$~direction and $h^x$ is a uniform transverse field.

According to the adiabatic theorem, the system remains in its instantaneous ground state (and the final state thus corresponds to the desired ground state of $\Hcost$), provided the Hamiltonian changes sufficiently slowly such that the adiabatic condition holds~\cite{Jansen2007,Lidar2009,Amin2009,Cheung2011}:
\begin{align}
    \label{eq:adiabatictheorem}
    \sweepduration \gg \hbar \max_{s \in [0, 1]} \frac{\abs[\big]{\braketx[\big]{n(s)}{\diff \hat{H}(s) / \diff s}{0(s)}}}{\Delta E_{n0}^2(s)} \,, \ \forall n \ge 1 \,.
\end{align}
Here, $\Delta E_{n0}(s)$ denotes the energy difference between the $n$-th instantaneous excited state $\ketx{n(s)}$ and the instantaneous (non-degenerate) ground state $\ketx{0(s)}$.

It should be noted that adiabaticity may not strictly be necessary for a successful annealing sweep.
In fact, the search for optimized (non-adiabatic) annealing schedules and tighter bounds for the minimal quantum annealing time is an active field of research, see for example Refs.~\cite{Cepaite2023,GarciaPintos2024, Bottarelli2024}.
For our purposes, we will nonetheless resort to the adiabatic condition in \cref{eq:adiabatictheorem} to assess the performance of quantum annealing for solving PUBO or QUBO problems.

\subsection{Estimating PUBO speedup for quantum annealing}\label{subsec:PUBOspeedup}

An order-of-magnitude estimate of the time required for a successful adiabatic sweep can be obtained from the right-hand side of the adiabatic condition in \cref{eq:adiabatictheorem}.
For our purposes, we employ the following simplified sufficient condition for adiabaticity:
\begin{align}
    \label{eq:adiabaticity}
    \sweepduration \gg \hbar \frac{V}{\Delta E^2} \equiv T \,.
\end{align}
Here, $\Delta E = \min_{s \in [0, 1]} \Delta E_{10}(s)$ is the minimum energy gap between the instantaneous first excited state and the instantaneous ground state, while the matrix elements in the numerator of \cref{eq:adiabatictheorem} are incorporated in the quantity $V = \max_{s \in [0, 1]} \mathinner{\abs{\braketx{1(s)}{\diff \hat{H}(s) / \diff s}{0(s)}}}$ (if the first excited state is degenerate, we take the maximum over the degenerate state manifold).
Throughout this work, we will refer to the right-hand side of \cref{eq:adiabaticity} as the adiabaticity time~$\annealingtime$.

In this study, we are primarily interested in hard problems where the minimum gap decreases exponentially with the problem size~$N$, 
\begin{align}
    \label{eq:scalinggap}
    \minimumgap = \mathinner{\epsilon} \etothe{-\alpha N} \,,
\end{align}
with characteristic energy~$\epsilon$ and exponent~$\alpha > 0$.
Such an exponential closing of the gap is a signature of a first-order quantum phase transition~\cite{Altshuler2010}.
By contrast, the quantity~$V$ in the numerator of \cref{eq:adiabaticity} scales at most proportional to the norm of the Hamiltonian, i.e., as $\mathcal{O}(N)$ for extensive Hamiltonians.
Consequently, the adiabaticity time is dominated by the behavior of the minimum gap and grows exponentially with the problem size as $\annealingtime \propto \etothe{2 \alpha N}$, up to polynomial corrections.
In our analysis, we will therefore use $\Delta E$ as the main figure of merit for the hardness of solving an optimization problem with quantum annealing.

When comparing the relative annealing performance for solving a problem posed in PUBO form versus its equivalent QUBO reduction, it is important to keep in mind that the energy scale of the Hamiltonian is typically different for the two problem types: the characteristic energy scale of the cost Hamiltonian encoding a PUBO problem is set by the three-body interaction strength~$J^{(3)}$, while for a QUBO problem it is determined by the two-body interaction strength~$J^{(2)}$.
Typically, three-body interactions are slower than two-body interactions, but the precise overhead depends on the hardware platform and implementation details (see also \cref{sec:3body}). 
One can keep track of this by expressing the quantities in the relations~\labelcref{eq:adiabaticity,eq:scalinggap} in terms of these scales, i.e., by defining the dimensionless quantities 
$\tilde{V}_P =  V_P/ J^{(3)}$ and $\tilde{V}_Q =  V_Q / J^{(2)}$, as well as  
$\tilde{\epsilon}_P = \epsilon_P / J^{(3)}$ and 
$\tilde{\epsilon}_Q = \epsilon_Q / J^{(2)}$, where $P$ and $Q$ denote PUBO and QUBO encoding, respectively. 
One can then estimate the potential PUBO speedup with respect to the equivalent QUBO formulation of a problem as the ratio of the corresponding adiabaticity times,   
\begin{equation}
\label{eq:pubospeedup}
    \frac{\annealingtime_Q}{\annealingtime_P} 
    = \frac{J^{(3)}}{J^{(2)}} \frac{\tilde{V}_Q}{\tilde{V}_P} \left( \frac{\tilde{\epsilon}_P}{\tilde{\epsilon}_Q} \right)^2 \etothe{2 (\alpha_Q - \alpha_P) N} \,.
\end{equation}

As \cref{eq:pubospeedup} illustrates, a potential PUBO speedup for solving exponentially hard problems with quantum annealing ($\annealingtime_Q / \annealingtime_P > 1$) is primarily determined by the size of the minimum gap and its scaling with the problem size, assuming at most a polynomial scaling for the various prefactors.
Specifically, under the above assumption and if $\alpha_P < \alpha_Q$, there is always an $N$ above which PUBO outperforms QUBO.
Even if PUBO and QUBO scale similarly with the system size ($\alpha_P \approx \alpha_Q$), assuming a not too small ratio $\tilde{V}_Q / \tilde{V}_P$, PUBO may be advantageous if the minimum gap has a larger offset, $\epsilon_P > \epsilon_Q$, which in spite of the typically slower three-body interaction rate is often the case due to the higher resource efficiency of the PUBO encoding.
In our numerical study of PUBO versus QUBO formulations of 3-SAT problems in \cref{sec:numerics}, we will encounter examples of both behaviors, $\alpha_P \approx \alpha_Q$ and $\alpha_P < \alpha_Q$, see \cref{fig:gapscaling:a,fig:gapscaling:b}, respectively.
Together with estimates for the overhead of the three-body relative to the two-body interaction, this allows us to estimate the potential for PUBO speedups for the investigated problem classes, see \cref{sec:3body}.

\subsection{Analog versus digital quantum annealing\label{sec:annealing:digital}}

Quantum annealing has originally been conceptualized for analog platforms, where the relative strength of the cost and driving Hamiltonian can be tuned continuously.
With the increasing availability of digital quantum computers, variants of quantum annealing adapted to gate-based architectures have become popular.
Digital annealers are particularly interesting in the light of direct implementations of PUBO as they allow one to synthesize the required higher-order interactions from universal sets of two- and single-qubit gates, whereas engineering native higher-order interactions in analog annealers is more challenging (see discussion in \cref{sec:3body:realization}).
For this reason, we briefly introduce a digital version of quantum annealing in this section.
Our analysis of PUBO versus QUBO performance applies to both analog and digital protocols alike as long as they are based on adiabatic evolution (in distinction to certain quantum-annealing-inspired algorithms like the Quantum Approximate Optimization Algorithm (QAOA)~\cite{Farhi2014}, where optimal performance requires schedules with large angles corresponding to evolution times outside the adiabatic regime).

Digital quantum annealing (or Trotterized quantum annealing) is accomplished via a Trotterization of the time evolution under the Hamiltonian in \cref{eq:annealingprotocol}.
That is, for a grid of times $0 = t_0 < t_1 < \dots < t_M = \sweepduration$ with uniform step size $\Delta t = \sweepduration / M$, the time evolution operator can be approximated as~\cite{Sack2021}
\begin{multline}
    \hat{\mathcal{T}} \etothe{-i \sweepduration \int_0^{1} \diff s^\prime \, \hat{H}(s^\prime) / \hbar} \\
    = \prod_{m = 1}^{M} \etothe{-i \Delta t (1 - s_{m}) \Hdrive / \hbar} \etothe{-i \Delta t s_{m} \Hcost / \hbar} + \mathcal{O}(\Delta t) \,,
\end{multline}
where $\hat{\mathcal{T}}$ is the time-ordering operator and $s_{m} = t_m / \sweepduration$.
Note that the local error for each Trotter step is $\mathcal{O}(\Delta t^2)$, such that the accumulated global error scales as $\mathcal{O}(\Delta t)$, which may be improved through higher-order splitting methods~\cite{Blanes2024}.
Since all terms within the cost Hamiltonian commute with each other, the corresponding matrix exponential can be exactly decomposed into a series of single-qubit $Z$~rotations and two-qubit $ZZ$~rotations. Similarly, the matrix exponential of the driving Hamiltonian can be decomposed into distinct single-qubit $X$~rotations. Each two-qubit rotation may then further be compiled into the native gate set of the respective architecture.

Compared to analog quantum annealing, this procedure incurs additional errors due to the discretization. Furthermore, since each time step $\Delta t$ needs to be decomposed into the native gate set, there is in general a polynomial overhead in the number of gates. 
For sufficiently small Trotter steps, however, the limiting factors to the algorithm's performance are the same as those of the continuous protocol.
In this situation, the adiabaticity time estimated from the minimum energy gap of the analog quantum annealing protocol acts as a proxy for the number of discrete time steps and thus the circuit depth required to reach the ground state of the cost Hamiltonian with the discrete protocol.
Unless stated otherwise, we will generally consider analog quantum annealing throughout this work.

\section{PUBO problems and their resource requirements\label{sec:pubo}}

In this section,
we discuss two instructive examples of optimization problems that can naturally be formulated as PUBO.
We start in \cref{subsec:toy_model} by illustrating the advantages of the PUBO formulation in terms of spatial resources, i.e., number of required qubits, for the simple toy model of minimizing a polynomial function.
In \cref{subsec:3sat}, we dive in some detail into the paradigmatic 3-SAT problem, which is the model we use for our numerical analysis in the rest of this work.

\Cref{app:other_PUBO} briefly presents a collection of further examples (the $p$-spin model, hypergraph coloring, NAE-k-SAT, and the traveling salesperson problem with time windows), illustrating the ubiquity of PUBO problems.

\subsection{Toy model: minimization of a polynomial function\label{subsec:toy_model}}

As a first simple example, we consider a continuous polynomial $f: \halfopeninterval{a,b} \to \mathbb{R}$ as an objective function to be minimized.
Without loss of generality, we choose the boundaries of the half-open interval as $a = 0$ and $b = 2^n$ for a convenient encoding of the continuous variable into bits.
We can then represent any continuous variable~$x$ approximately as a bitstring $(x_1, \dots, x_{n + m})$ via the mapping
\begin{align} 
    x \approx \sum_{i=1}^n 2^{n-i} x_i + \sum_{i=1}^m 2^{-i}x_{n+i} \,, \label{eq:cont_bit_encoding}
\end{align}
where the first sum corresponds to the integer part and the second sum to the decimal part with $m$-digit resolution. Note that negative numbers can be handled in a similar way by introducing an additional Boolean variable $x_0$, encoding the sign of $x$, and multiplying \cref{eq:cont_bit_encoding} by $(2x_0 -1)$~\cite{Stein2023}. 

As a concrete case, take the task to find the minimum of the cubic polynomial $f(x) = x^3 + x$ in the domain $\halfopeninterval{0,2^2}$. After inserting the encoding in \cref{eq:cont_bit_encoding} with a resolution of $m=1$, we obtain
\begin{align}
\label{eq:cont_pubo}
\begin{split}
    f(x_1,x_2,x_3) &= 4 x_1 + 2 x_2  + \frac{5}{8} x_3 + 6 x_1 x_2 \\
    &\hphantom{{}=} + \frac{15}{2} x_1 x_3  + \frac{9}{4} x_2 x_3 + 6 x_1 x_2 x_3 \,.
\end{split}
\end{align}
Due to the term $\propto x_1 x_2 x_3$, this optimization problem is of third-order PUBO form. The trivial solution in this case is given by $(x_1, x_2, x_3) = (0,0,0)$.

\Cref{eq:cont_pubo} can be reduced to a QUBO problem by introducing an additional slack variable $y = x_2 x_3$. An appropriate penalty term $\propto (3y + x_2 x_3 - 2x_2y - 2x_3 y) $  needs to be added to the optimization problem in order to enforce consistency.
While in this specific case the QUBO formulation requires only a single extra qubit compared to the equivalent PUBO formulation, the resource savings can get significant for larger and more complex problems (see below).
Furthermore, the optimal introduction of slack variables (i.e., the assignment which requires the least amount of additional bits) is itself known to be an \complexityclass{NP}-hard problem~\cite{Boros2002}. 
Finally, the constraint that enforces consistency of the ancillary variables needs to be sufficiently strong such that other terms cannot overpower it. Since the maximal energy scales available in a given quantum machine are limited, this means in practice reducing the relative strength of the other terms, and thus incurring proportionally longer sweep times or deteriorating the adiabaticity of the sweep.
Finding the minimum penalty strength that enforces all constraints is in general \complexityclass{NP}-hard, but efficient approximate schemes with significant performance gains over default choices have been proposed~\cite{Alessandroni2023}.

\subsection{Boolean satisfiability problem\label{subsec:3sat}}

Given a Boolean formula, the satisfiability problem (SAT) poses the decision problem whether there is an assignment of truth values such that the formula evaluates to \textit{true}.
In the case of $k$-SAT, the formula takes the form of a conjunction of $M$ logical clauses $C_m$, 
\begin{align}
    C_{1} \land C_{2}\land ...\land C_M \,,
\end{align}
with each clause being composed of a disjunction of $k$ literals chosen from a set of $N$ Boolean variables $x_i$ and their negations $\lnot{x_i}$.
An example clause for $k=3$ is 
\begin{align}
    C_{m}(x_{i}, x_{j}, x_{l})= x_{i}\lor \lnot{x_{j}} \lor\lnot{x_{\ell}} \,.
\end{align}
Every SAT problem can be rewritten as a $k$-SAT problem using the rules of Boolean algebra.
It is well known that $k$-SAT for $k\leq 2$ belongs to the complexity class~$\complexityclass{P}$, i.e., it is solvable in polynomial time on a classical computer~\cite{Aspvall1982}.
In contrast, 3-SAT was the first decision problem proven to be \complexityclass{NP}-complete (Cook--Levin theorem~\cite{Cook1971, Karp1972}). If $\complexityclass{P} \neq \complexityclass{NP}$, no classical algorithm could solve 3-SAT in polynomial time. Additionally, it is widely believed that problems in \complexityclass{NP}-complete are not contained in $\complexityclass{BQP}$ \cite{Aaronson2010}, and could therefore not be solved in polynomial time on a quantum computer either.
Nevertheless, even if an exponential quantum advantage may be out of reach, one can still hope for a practical speedup through quantum annealing~\cite{Heim2015, Albash2018, Willsch2022, Slongo2023, Bernaschi2024}.

A phase transition for 3-SAT problems is known to occur asymptotically at a critical ratio of clauses to variables of $(M/N)_{\mathrm{crit}} \approx \num{4.24}$~\cite{Mezard2002}. Around this critical point, one is likely to encounter hard-to-solve problems with unique solutions. Below the phase transition point, most problem instances are easy to solve with many solutions, while above the transition most instances become unsatisfiable.
Note that although the probability of finding instances with unique solutions below the phase transition point decreases exponentially, it is well-known that those rare instances (with fewer clauses and unique solutions) tend to be the hardest ones to solve~\cite{Znidaric2005}.
Our results in \cref{sec:numerics} underline this fact by comparing the minimum energy gaps of instances constructed randomly by different generators with different ratios of clauses to variables.

\subsubsection{3-SAT as PUBO}\label{subsubsec:3satpubo}

3-SAT can be reformulated in various ways as a PUBO or QUBO problem in order to make it suitable for quantum annealing.
A straightforward encoding consists in a PUBO problem of degree three: each clause is cast into the form of an energy term penalizing any choice of truth values that does not satisfy the clause. For example, the clause $C_{m}= x_{i}\lor \lnot{x_{j}} \lor\lnot{x_{\ell}}$ from above can be translated to $(1-x_i)x_{j}x_{\ell}$, which only evaluates to zero if at least one variable satisfies the corresponding clause. Adding up these terms for each clause and translating the binary variables to quantum spin variables via $x_{i} = 1/2 - \hat{s}^{z}_{i}$, leads to the general three-body spin Hamiltonian in \cref{eq:3SAT_PUBO}.
Note that, in this specific PUBO formulation, the three-body and two-body interaction strengths $J^{(3)}$ and $J^{(2)}$ as well as the local biases~$\vec{h}^z$ are all of the same order of magnitude or zero.
The ground state of the cost Hamiltonian then encodes the solution of the original 3-SAT problem (more precisely, it solves MAX-SAT, i.e., finding the solution that satisfies most clauses).
This encoding requires $N$ spins (qubits) for the $N$ Boolean variables in the original problem.

\subsubsection{3-SAT as QUBO via slack variables}\label{subsubsec:3satquboreduc}

Currently available quantum annealing hardware (e.g., the D-Wave machine~\cite{dwave}) natively only supports two-body interactions. Thus, one needs to find an equivalent encoding as a QUBO problem to solve 3-SAT on such hardware. A resource-efficient way to achieve this is to start with the PUBO formulation of 3-SAT given by the general Hamiltonian in \cref{eq:3SAT_PUBO} and to introduce slack variables that reduce all three-body interactions to quadratic and local terms (see \cref{subsec:toy_model}).
Since there are at most $M$ (number of clauses) unique three-body terms in the Hamiltonian, at most $M$ additional qubits are required by this QUBO reduction.
For hard problem instances around the phase transition, which we are most interested in here, this amounts to at most around $4.24 \, N$ additional qubits. 

In \cref{app:QUBO_reducs}, we discuss two other approaches to formulate 3-SAT as a QUBO problem. However, the direct approach of introducing slack variables is the most efficient in terms of qubit requirements and will be considered for the rest of this work.

\section{Numerical results}\label{sec:numerics}

\begin{figure*}[ht]
    \includegraphics[width=\linewidth]{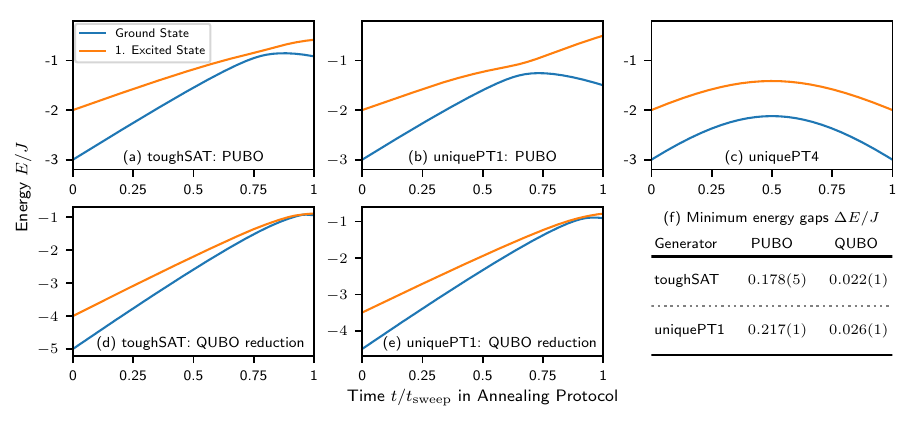}%
    \subfloat{\label{fig:typicalgaps:a}}%
    \subfloat{\label{fig:typicalgaps:b}}%
    \subfloat{\label{fig:typicalgaps:c}}%
    \subfloat{\label{fig:typicalgaps:d}}%
    \subfloat{\label{fig:typicalgaps:e}}%
    \subfloat{\label{fig:typicalgaps:f}}%
    \caption{\label{fig:typicalgaps}%
    Typical gap profiles of PUBO and QUBO formulations of randomly generated 3-SAT instances.
    The plots show the energies of the ground state (blue) and first excited state (orange) as a function of time during an annealing sweep according to \cref{eq:annealingprotocol} with driving strength $h_x$ equal to the characteristic energy scale $J$ of the cost Hamiltonian.
    The upper panels (a) and (b) depict the native PUBO formulations (cost Hamiltonian with up to three-body interactions, $J = \max_{ijk} |J_{ijk}^{(3)}|$), while the lower panels (d) and (e) show the associated QUBO reductions (cost Hamiltonian with at most two-body interactions, $J = \max_{ij} |J_{ij}^{(2)}|$). Note that two-body interactions are typically much stronger than three-body interactions. A direct decomposition of a 3-qubit unitary gate gives a ratio of $J^{(3)}/J^{(2)} \approx 1/4$ (see \cref{sec:3body:realization}).
    The 3-SAT instances comprise $N = 6$ classical variables and are generated by
    the \algorithm{toughSAT} generator with $M = \round(4.24 \, N) = 25$ clauses [panels (a) and (d)], 
    the \algorithm{uniquePT1} generator with $M = N + 6$ clauses [panels (b) and (e)], and
    the \algorithm{uniquePT4} generator with $M/N = 4$ clauses [panel~(c)].
    The QUBO reductions of the instances in (a) and (b) require four and three additional slack variables (ancillary qubits), respectively.
    For instances generated by \algorithm{uniquePT4} (c), no gap closing occurs since the generated instances are trivial (all three-body and two-body interactions cancel, $J = \max_{i} |h_i^z|$).
    The table~(f) shows the mean minimum energy gap and its standard error of the mean for an ensemble of 200 instances corresponding to the scenarios in (a), (b) and (d), (e).
    }
\end{figure*}

In this section, we numerically test the quantum annealing performance of PUBO versus QUBO formulations for different instances of the 3-SAT problem (see \cref{subsec:3sat}) using exact diagonalization.
We focus our analysis on the behavior of the minimum energy gap~$\minimumgap$ encountered during the annealing protocol, which is the key quantity that mainly determines the annealing performance, as explained in \cref{sec:annealing}.

\subsection{Methods}

In what follows, we briefly outline the framework and key algorithms employed in our numerical study.
Throughout this section, we consider the annealing schedule given in \cref{eq:annealingprotocol}, where the cost Hamiltonian is ramped up linearly, while the strength of a homogeneous transverse field is linearly decreased.

\subsubsection{Characteristic energy scales\label{subsubsec:energyscales}}

In our numerics, we express the cost Hamiltonian~$\Hcost$ in units of the global energy scale~$J$, which we identify with a characteristic energy of the physical device.
In principle, it is desirable for $J$ to be as large as possible in order to speed up the absolute computing time.
In practice, however, $J$ is limited by the maximum physical coupling rates the device can operate with.
Consequently, a global rescaling of the Hamiltonian is required to ensure that none of the terms in \cref{eq:3SAT_PUBO} exceed these limits.
On analog hardware, the physical couplings relevant for PUBO implementations up to third order include the maximum physical three-body and two-body interaction strengths as well as the maximum strength of local fields.
On digital devices, the physical time scales associated with these interactions correspond, respectively, to the three-, two-, and single-qubit gate times.
Since interactions of higher rank are generally slower compared to interactions of lower rank (see discussion in \cref{sec:3body:realization}), the former typically create a bottleneck: the global energy scale $J$ is limited by the maximum physical three-body interaction strength if three-body interactions are present (third-order PUBO) and by the maximum physical two-body interaction strength if only two-body interactions are present (QUBO), where we assume that local fields can be scaled sufficiently high to pose no restriction.
In particular, three-body interactions are the limiting factor for PUBO implementations of typical 3-SAT instances encountered in our small-scale numerical experiments, where the maximum ratio of three-body and two-body couplings is of order one (though this may be different for larger instances or problems with different PUBO structures).
It is therefore natural in our case to choose the characteristic energy $J = \max_{ijk}|J_{ijk}^{(3)}|$ as the global energy unit for third-order PUBO, while for QUBO we use $J = \max_{ij} |J_{ij}^{(2)}|$.%

Importantly, for QUBO reductions of native PUBO problems, $\Hcost$ incorporates both the problem Hamiltonian and penalty terms implementing consistency constraints. The two-body interactions can then be separated: $J_{ij}^{(2)} = [J_{ij}^{(2)}]_{\mathrm{problem}} + [J_{ij}^{(2)}]_{\mathrm{constraints}}$.
Typically, enforcing the constraints requires the penalty terms to be much larger than the QUBO terms of the actual problem Hamiltonian.
Since the characteristic energy scale $J = \max_{ij} |J_{ij}^{(2)}|$ is limited by the maximum physical interaction rate supported by the hardware, we end up with $[J_{ij}^{(2)}]_{\mathrm{problem}} / J \ll 1$ and $[J_{ij}^{(2)}]_{\mathrm{constraints}} / J \sim 1$, i.e., the part of the Hamiltonian encoding the actual problem is scaled down with respect to the penalty terms.
In \cref{subsubsec:penalty}, we investigate how the relative strength of the penalty terms influences the minimum energy gap and thus the adiabaticity time.

Unless stated otherwise, we choose the strength of the driving Hamiltonian in \cref{eq:annealingprotocol} to be the same as the characteristic energy scale of the cost Hamiltonian, $h^x = J$.
In \cref{subsec:drivingvscost}, we justify this choice and discuss the effects of varying $h^x$ on the annealing performance.

\subsubsection{3-SAT problem generators}

To generate instances of the 3-SAT problem, we adapt the three different generators \algorithm{toughSAT}, \algorithm{uniquePT1}, and \algorithm{uniquePT4} described in Ref.~\cite{Hsieh2021}.

The \algorithm{toughSAT} generator creates an arbitrary number~$M$ of 3-SAT clauses by randomly picking three out of $N$ variables and then randomly applying negations to them.
In our numerics, we set the ratio of clauses to variables close to the critical ratio $(M / N)_{\mathrm{crit}} \approx 4.24$, as we expect to find hard-to-solve instances with unique ground states in the vicinity of this point (see \cref{subsec:3sat}).
To avoid ground state degeneracies, we postselect the randomly generated instances for those with unique solutions.

In contrast, the generators \algorithm{uniquePT1} and \algorithm{uniquePT4} always create instances with known and unique solutions (although not necessarily hard to solve in the latter case). Both generators work by picking a solution randomly and then constructing appropriate 3-SAT clauses to make said solution unique. While \algorithm{uniquePT4} creates $M = 4N$ clauses to achieve that, \algorithm{uniquePT1} only needs $M = N + 6$ clauses. 
However, we find that \algorithm{uniquePT4} constructs clauses in such a way that all third-order and second-order terms in the PUBO cost function, i.e., three-body and two-body interactions in the cost Hamiltonian, cancel each other.
What remains is a simple cost Hamiltonian with only local terms of the form $\Hcost= -\sum_i h_i^z \hat{s}_i^z$, whose solution can easily be read off. Thus, \algorithm{uniquePT4} creates trivial 3-SAT problems, where the time required to solve them scales linearly in the problem size.

For our numerical analysis, we are primarily interested in hard-to-solve problem instances generated by \algorithm{toughSAT} and \algorithm{uniquePT1}, which exhibit an exponential closing of the minimum energy gap with increasing problem size. For comparison,
we also show the spectrum of a problem instance generated by \algorithm{uniquePT4}, see \cref{fig:typicalgaps:c}.

\subsubsection{PUBO-to-QUBO reduction\label{sec:quboreductionnumerics}}

To assess the annealing performance gains of PUBO over QUBO formulations when solving 3-SAT instances, we compare the minimum energy gap of the PUBO formulation with that of the equivalent QUBO reduction.
Finding an optimal QUBO reduction for a given PUBO problem is in general \complexityclass{NP}-hard.
For the small system sizes considered here, we employ the following brute-force algorithm: replace the pair of variables that appears most frequently in the third-order terms of the PUBO cost function by a slack variable (see \cref{subsec:toy_model}), then continue with the second-most-frequent pair of variables and so on, until all third-order terms have been replaced by quadratic ones with associated slack variables and constraints~\cite{Verma2021}.
For each newly introduced ancilla variable $y = x_i x_j$, a penalty term $\propto (3y + x_i x_j - 2x_i y - 2x_j y) $ needs to be added to the cost Hamiltonian (see \cref{subsec:toy_model}) and scaled by a suitable factor, thus ensuring that violating the penalty term in favor of the three-body interactions is never energetically favorable. This factor depends on the problem size and structure: in our numerics, we make the simple choice that for every three-body interaction term $a_{ijk} x_i x_j x_k$ that is to be reduced by the ancilla variable $x_{\mathrm{ancilla}} = x_i x_j$, we add $|a_{ijk}| + 1$ to the strength of the corresponding penalty term.
In \cref{subsec:drivingvscost}, we discuss how varying the strength of the penalty terms relative to the original cost Hamiltonian affects the overall annealing performance.

The QUBO formulation of \algorithm{toughSAT} instances usually requires more slack variables than \algorithm{uniquePT1} instances of the same size. This is due to the fact that the number of ancilla variables is bounded from above by the number of clauses in the problem instance. 
For both generators,
the increased resource requirements of the QUBO reductions generally lead to a smaller minimum gap and thus to a reduced annealing performance, as we will see below.

\subsection{Analysis of the minimum gap}

In what follows, we present our numerical results on the behavior of the minimum energy gap for randomly generated 3-SAT instances.
To begin with, we illustrate the increase of the minimum gap in the PUBO formulation of the 3-SAT problem compared to the equivalent QUBO reduction and discuss the correlation of the minimum gap with the number of three-body interaction terms in the cost Hamiltonian. We then compare the scaling of the minimum gap with system size for the equivalent PUBO and QUBO formulations.

\subsubsection{Gap profiles}\label{subsubsec:gapprofiles}

\Cref{fig:typicalgaps} illustrates typical gap profiles during an annealing sweep for specific instances with $N=6$ variables, generated by each of the three generators. The blue and yellow lines correspond, respectively, to the energies of the instantaneous ground state and first excited state during a sweep according to the annealing protocol in \cref{eq:annealingprotocol}.
While the \algorithm{toughSAT} and \algorithm{uniquePT1} instances exhibit significant gap closings, the energy gap of the \algorithm{uniquePT4} instance stays consistently large over the entire sweep, indicating that the generated instances are trivial to solve.
The gap profile of the depicted \algorithm{uniquePT4} instance is representative of every random instance created by this generator.

For the \algorithm{toughSAT} and \algorithm{uniquePT1} instances, we also plot the energies of the ground state and first excited state in the corresponding QUBO reduction.

In \cref{fig:typicalgaps:f}, we report the mean minimum energy gap and its standard error of the mean, obtained by averaging over a large number of samples.

It can be seen that for this small system size, \algorithm{uniquePT1} and \algorithm{toughSAT} produce on average comparable energy gaps (slightly larger for the former), even though \algorithm{uniquePT1} 
requires less clauses and thus less additional slack variables in the QUBO reduction. 
For larger systems, this is no longer true, as we will see in \cref{subsec:scaling}.

Importantly, the gaps in the QUBO reductions are significantly smaller compared to those in the corresponding PUBO formulations.
The larger
relative 
variance in the mean energy gaps of the QUBO reductions can be explained by the fact that not every generated 3-SAT instance requires the same amount of slack variables for its reduction to a QUBO problem.

Recall that when characterizing the relative performance of PUBO versus QUBO via the adiabaticity time, it is necessary to take into account the different physical energy scales set by the three-body and two-body interaction rates $J^{(3)}$ and $J^{(2)}$, respectively, see \cref{subsubsec:energyscales}.
We will come back to this point in \cref{sec:3body}.

\subsubsection{Absence of correlations between minimum gap and number of three-body terms}

\begin{figure}[ht]
    \raggedright
    (a)\\
    \includegraphics[width=\linewidth]{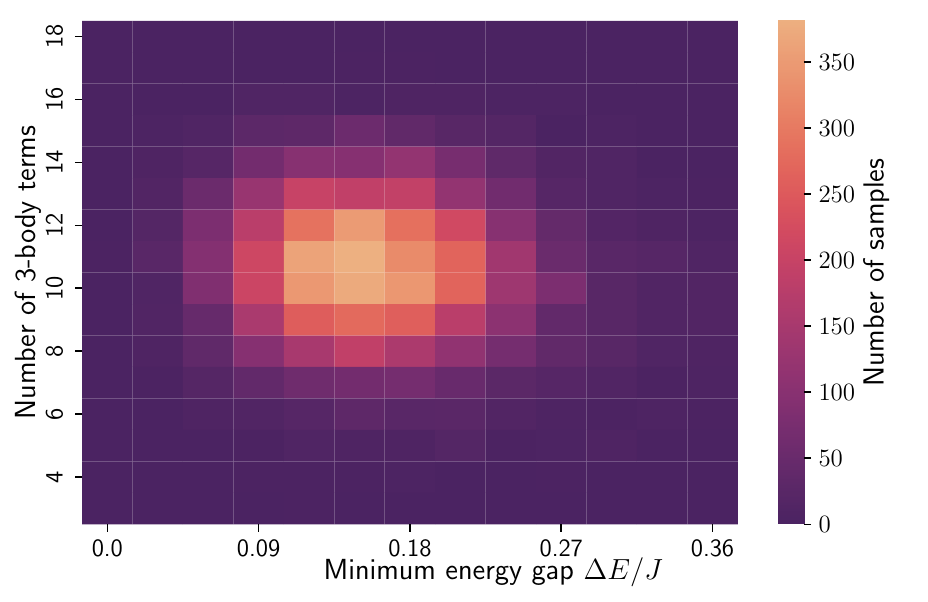}\\
    (b) \hspace{11em}
    (c) \\
    \includegraphics[width=0.49\linewidth]{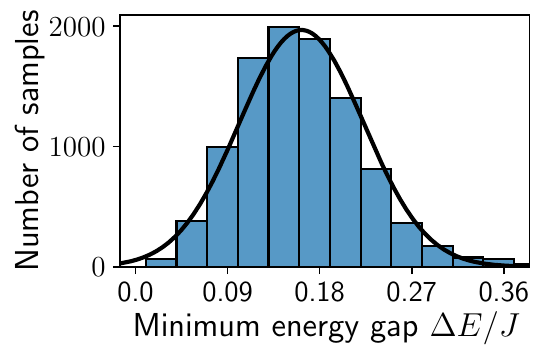}
    \includegraphics[width=0.49\linewidth]{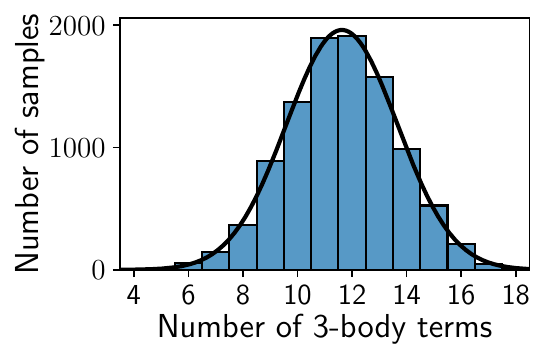}%
    \subfloat{\label{fig:corrgapterms:a}}%
    \subfloat{\label{fig:corrgapterms:b}}%
    \subfloat{\label{fig:corrgapterms:c}}%
    \caption{\label{fig:corrgapterms}%
    Distribution of the minimum energy gap and the number of third-order terms in the PUBO cost function (nonzero three-body interaction coefficients in the cost Hamiltonian).
    (a)~Two-dimensional histogram, computed for $\num{10000}$ random 3-SAT instances generated via \algorithm{toughSAT} with $N=6$ and $M = 25$.
    (b)--(c)~Marginal histograms of the minimum energy gap and the number of third-order terms, respectively, with fitted Gaussian distributions.
    No significant correlations can be discerned. 
    }
\end{figure}

One may wonder whether the minimum gap (and thus the hardness to solve the PUBO problem with quantum annealing) depends on the number of nonzero three-body interaction coefficients present in the cost Hamiltonian.
\Cref{fig:corrgapterms} investigates the correlations between these two quantities for $\num{10000}$ random instances generated via \algorithm{toughSAT} with $N=6$ and 
$M = \round(4.24 \, N) = 25$.
The samples for both the minimum gap and the number of three-body terms are broadly distributed around the mean, and are well approximated by a Gaussian (see marginal distributions in Fig.~\ref{fig:corrgapterms}). 
The number of three-body terms spreads in the range $\numrange{4}{19}$.
We do not find any significant correlation between the minimum gap and the number of three-body terms.

\subsubsection{Scaling of the minimum gap versus system size}
\label{subsec:scaling}

\begin{figure}
    \centering
    \includegraphics[width=\columnwidth]{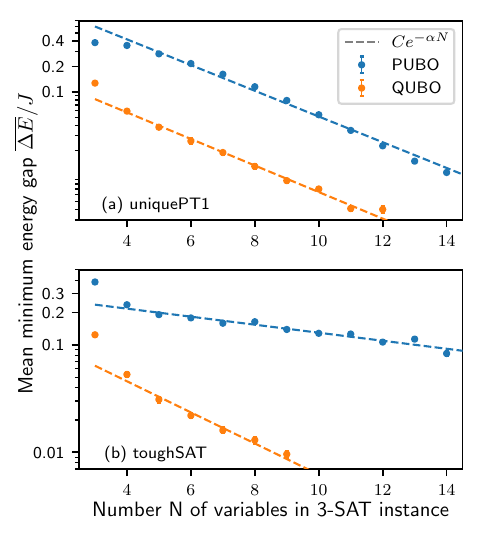}%
    \subfloat{\label{fig:gapscaling:a}}%
    \subfloat{\label{fig:gapscaling:b}}%
    \caption{\label{fig:gapscaling}%
    Scaling of mean minimum energy gap~$\ensembleaverage{\minimumgap}$ with the number of Boolean variables~$N$ for ensembles of 200 randomly generated 3-SAT instances.
    The error bars show the statistical standard error of the mean, 
    which is smaller than the marker size for most data points.
    The dashed lines show exponential fits to the data 
    (ignoring $N=3$) 
    according to $\ensembleaverage{\minimumgap} = \mathinner{\epsilon} \etothe{-\alpha N}$ with fit parameters $\epsilon$ and $\alpha$ given in \cref{tab:minimalgap_fit}.
    For problems generated with \algorithm{uniquePT1} (a), the gap closes approximately at the same rate in both the PUBO and QUBO formulation, while for \algorithm{toughSAT} (b), the QUBO formulation exhibits a faster gap closing with increasing $N$.
    In all cases, the minimum gap relative to the characteristic energy scales is larger for the PUBO  formulation ($J \sim J^{(3)}$) than for the corresponding QUBO reduction ($J \sim J^{(2)}$).
    }
\end{figure}

\begin{table}[ht]
    \centering
    \sisetup{round-mode=uncertainty,round-precision=1}
    \begin{tabularx}{\linewidth}{YYYYY}
    \toprule
                   & \multicolumn{2}{c}{PUBO} & \multicolumn{2}{c}{QUBO} \\
    \cmidrule(lr){2-3}\cmidrule(lr){4-5}
         generator & $\epsilon_P / J$ & $\alpha_P$ & $\epsilon_Q / J$ & $\alpha_Q$ \\
    \midrule
         \algorithm{toughSAT}   & $\num{0.306\pm 0.008}$  & $\num{0.086 \pm 0.004}$ & $\num{0.174 \pm 0.019}$ & $\num{0.333 \pm 0.017}$ \\
         \algorithm{uniquePT1}  & $\num{1.721 \pm 0.013}$ & $\num{0.352 \pm 0.001}$ & $\num{0.244 \pm 0.013}$ & $\num{0.364 \pm 0.008}$ \\
    \bottomrule
    \end{tabularx}
    \caption{\label{tab:minimalgap_fit}%
    Parameters of the fits shown in \cref{fig:gapscaling}, characterizing the exponential scaling of the mean minimum energy gap $\ensembleaverage{\minimumgap} = \mathinner{\epsilon} \etothe{-\alpha N}$ for randomly generated 3-SAT instances.
    The uncertainty of the fit parameters corresponds to the standard error of the mean for $200$ realizations.
    }
\end{table}

As explained in \cref{sec:annealing}, a key characteristics for predicting the performance of a quantum annealer is how the minimum energy gap encountered during the sweep scales with the problem size.
In \cref{fig:gapscaling}, we analyze the behavior of the minimum gap as a function of the number of variables for 3-SAT problems in their PUBO form as well as in their equivalent QUBO reduction (see \cref{sec:quboreductionnumerics}).
The random 3-SAT instances are generated by the \algorithm{toughSAT} and \algorithm{uniquePT1} generators and averaged over 200 realizations.

In the investigated range of problem sizes~$N$, we find the mean minimum gap $\ensembleaverage{\minimumgap}$ in all cases to close exponentially according to $\ensembleaverage{\minimumgap} = \mathinner{\epsilon} \etothe{-\alpha N}$, as is to be expected for \algorithm{NP}-complete problems like 3-SAT. 
The parameters of the best fits are given in \cref{tab:minimalgap_fit}. 
\algorithm{toughSAT} exhibits a slower scaling in the PUBO formulation than \algorithm{uniquePT1}, indicating that the former generates on average easier (but still exponentially hard) instances.
Notably, for ensembles generated by \algorithm{toughSAT} [\cref{fig:gapscaling:b}], the gap closing in the PUBO formulation is significantly slower than in the QUBO formulation. 
Although the precise scaling may depend on the choice of the ratio of driving versus cost Hamiltonian as well as the strength of the penalty terms in the QUBO reduction,
we argue in the following sections that the employed parameters are chosen close to optimal for the problem sizes investigated. 
This scaling thus suggests that for sufficiently large problems the PUBO encoding will always be advantageous.
In the case of \algorithm{uniquePT1} [\cref{fig:gapscaling:a}], the gap closes approximately at the same rate for both PUBO and QUBO (slightly faster in the QUBO reduction).
According to \cref{eq:pubospeedup}, a potential PUBO speedup then depends on the relative strength of the three-body with respect to the two-body interaction, which we analyze in more detail in \cref{sec:3body}.

The numerically determined exponential scaling of the minimum energy gap also allows us to estimate the scaling of the adiabaticity time~$T$ as a figure of merit for the required time to solve the optimization problem with quantum annealing.
Assuming that the matrix elements contributing to the quantity~$V$ in \cref{eq:adiabaticity} scale at most polynomially with the system size~$N$, the scaling of the adiabaticity time is dominated by the behavior of the mean minimum energy gap, $T \propto \smash{\ensembleaverage{\minimumgap}}^{-2}$.
For PUBO formulations of \algorithm{toughSAT} instances, we then find that the adiabaticity time scales on average as $T \propto \num{1.19}^N$.
It is tempting to compare this number with classical solvers: the Unique 3-SAT problem, i.e., the promise version of 3-SAT, where each instance has either exactly one or no solution, can be solved classically in time $\propto 1.307^N$~\cite{Hansen2019}.
Remarkably, the adiabaticity time for \algorithm{toughSAT} instances in PUBO form grows more slowly with the system size.
Conversely, for the harder 3-SAT instances generated by \algorithm{uniquePT1}, we find a scaling of the adiabaticity time as $T \propto \num{2.02}^N$ for the PUBO formulation, which is close to the $2^N$ scaling of the classical brute-force approach.
Although the adiabaticity time is only a crude approximation of the actual runtime required by quantum annealing and may not reflect accurately the true computational complexity, our findings raise the possibility that, for certain problem families, quantum annealing with a PUBO formulation of 3-SAT could be competitive with or even surpass classical performance.

\subsubsection{Dependence of the minimum energy gap on the QUBO penalty strength\label{subsubsec:penalty}}

\begin{figure}
    \includegraphics{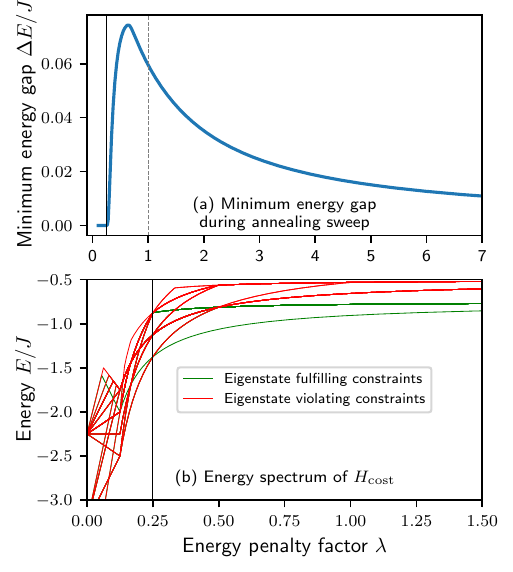}%
    \subfloat{\label{fig:penaltyscaling:a}}%
    \subfloat{\label{fig:penaltyscaling:b}}%
    \caption{\label{fig:penaltyscaling}%
    Dependence of the minimum energy gap on the strength of the penalty terms in the QUBO reduction of a random 3-SAT instance generated by \algorithm{uniquePT1} with $N = 6$ variables.
    (a)~Minimum energy gap $\Delta E / J$ ($J =\max_{ij} |J_{ij}^{(2)}|$) encountered during the annealing sweep as a function of the penalty strength~$\lambda$ in the cost Hamiltonian $\Hcost = \hat{H}_{\mathrm{problem}} + \lambda \hat{H}_{\mathrm{constraints}}$.
    The gray dashed line shows the value $\Delta E / J \approx 0.06$ corresponding to the default choice $\lambda = 1$ used throughout this work.
    In this example, the optimum $\Delta E / J \approx 0.074$ is reached at the slightly lower value $\lambda \approx \num{0.64}$.
    (b)~Energy spectrum of the cost Hamiltonian~$\Hcost$ for the lowest 30 eigenvalues, with most of them being degenerate.
    Levels whose associated eigenstates fulfill all (violate any) of the constraints are colored in green (red).
    The solid black lines in the panels (a) and (b) mark the location where the ground state becomes degenerate and no longer fulfills the consistency constraints, causing the minimum gap to vanish ($\lambda = \num{0.25}$ in this example).
    }
\end{figure}

Throughout our numerical analysis above, we have followed the heuristic described in \cref{sec:quboreductionnumerics} to determine the strength of the energy penalty terms in the QUBO reduction of a PUBO problem such that the solution satisfies all constraints. In this section, we investigate how varying this strength affects the minimum energy gap.

To this end, we separate the cost Hamiltonian as
\begin{equation}
\label{eq:penaltyhamiltonian}
    \Hcost = \hat{H}_{\mathrm{problem}} + \lambda \hat{H}_{\mathrm{constraints}}
\end{equation}
into the part encoding the actual problem, $\hat{H}_{\mathrm{problem}}$, and the part ensuring the fulfillment of consistency constraints in the QUBO reduction, $\hat{H}_{\mathrm{constraints}}$. 
Recall that the latter originates from penalty terms of the form $3y + x_i x_j - 2x_iy - 2x_j y$, one for each cubic term $\propto x_i x_j x_k$ that is to be reduced by an ancilla variable $y = x_i x_j$ in the PUBO cost function.
To tune the strength of the energy penalty, we have introduced a factor~$\lambda \ge 0$ in \cref{eq:penaltyhamiltonian}, which globally scales the strength of $\hat{H}_{\mathrm{constraints}}$ relative to $\hat{H}_{\mathrm{problem}}$ (while preserving the relative weights of the individual terms in $\hat{H}_{\mathrm{constraints}}$).
For each choice of
$\lambda$, the cost Hamiltonian is normalized as before such that the characteristic energy scale is $J = \max_{ij} |J_{ij}^{(2)}|$ with $J_{ij}^{(2)}(\lambda) = [J_{ij}^{(2)}]_{\mathrm{problem}} + \lambda [J_{ij}^{(2)}]_{\mathrm{constraints}}$, see \cref{subsubsec:energyscales}.
The value $\lambda = 1$ corresponds to the configuration used in the previous sections.

As a representative example, we consider a random 3-SAT instance generated by \algorithm{uniquePT1} with $N = 6$ variables.
\Cref{fig:penaltyscaling:a} shows the minimum energy gap as a function of the penalty strength~$\lambda$.
In this specific example, the largest value of the minimum energy gap, $\Delta E / J \approx 0.074$, is obtained for $\lambda \approx \num{0.64}$.
For comparison, our default choice, $\lambda = 1$, yields a gap that is about $\SI{19}{\percent}$ smaller, $\Delta E / J \approx 0.06$.
Higher values of $\lambda$ further reduce the minimum gap since the relative strength of $\hat{H}_{\mathrm{problem}}$ with respect to $\hat{H}_{\mathrm{constraints}}$ decreases.
Tuning $\lambda$ in the opposite direction below the optimal value leads to a rapid decrease of the gap until it reaches zero where the ground state becomes degenerate and changes to a state that no longer fulfills the consistency constraints.
In the example of \cref{fig:penaltyscaling:a}, the point where the solution becomes unfeasible is reached at $\lambda = \num{0.25}$, but for other problem instances, this can occur at larger values, e.g., $\lambda = \num{0.5}$, in which case the optimum value lies even closer to our default choice $\lambda = 1$.

To gain a deeper understanding of this behavior, we study the energy spectrum of the cost Hamiltonian~$\Hcost$ as a function of $\lambda$, shown in \cref{fig:penaltyscaling:b} for the lowest 30 eigenvalues (all levels except for the unique solution are degenerate).
For large enough values of the penalty parameter~$\lambda$, all constraint-violating eigenstates (red color) are shifted up in energy, so that both the ground state and the first excited state fulfill all constraints (highlighted in green). Normalizing $\Hcost$
to a fixed energy scale~$J$
reduces the energies (and thus the gap) of these two states by the factor $\propto \lambda^{-1}$.
Consequently, it is counterproductive to choose the energy penalty too strong.
Conversely, $\lambda$ should not be chosen too small either: the largest minimum energy gap occurs close to the value of $\lambda$ for which the first excited penalty-satisfying state and the first excited penalty-violating state become degenerate [$\lambda = 0.5$ in \cref{fig:penaltyscaling:b}].
Reducing $\lambda$ further brings the first penalty-violating state closer to the ground state, resulting in a sharp decrease of the minimum gap.

To summarize, our example demonstrates that there is in general potential for optimization of the minimum energy gap in the QUBO reduction of a native PUBO problem by tuning the strength of the penalty terms.
However, finding the optimal penalty strength generally requires solving the problem itself, which is prohibitively expensive for large exponentially hard problems.
For our scaling analysis of the minimum gap, we have thus chosen the penalty strength in a problem-agnostic way as described in \cref{subsec:scaling}.
This heuristic choice is simple, incurs only moderate performance overhead, and reliably produces feasible solutions.
Further optimizations may be achieved through more sophisticated schemes to approximate the optimal penalty strength~\cite{Alessandroni2023}, but such improvements would have only minor effects on the results of our PUBO versus QUBO analysis, as in our scenario the simple approach is already close to optimal.

\subsubsection{Influence of the driving strength on the annealing performance\label{subsec:drivingvscost}}

\begin{figure}
    \centering
    \includegraphics[width=\linewidth]{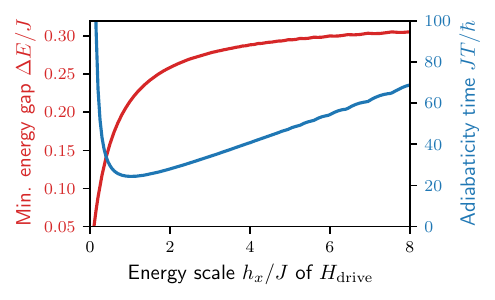}
    \caption{\label{fig:cost_to_driving_ratio}%
    Influence of the driving strength on the minimum energy gap and the adiabaticity time for a linear annealing schedule $(1-s)h_x \Hdrive^\prime + s J \Hcost^\prime$.
    The cost Hamiltonian $\Hcost = J \Hcost^\prime$ corresponds to the PUBO encoding of a random 3-SAT instance generated by \algorithm{uniquePT1} with $N = 6$ variables and has characteristic energy scale $J = \max_{ijk} |J_{ijk}^{(3)}|$.
    The driving Hamiltonian~$\Hdrive = h^x \Hdrive^\prime = - h^x \sum_i \hat{\sigma}_i^x$ represents a uniform transverse field of strength~$h^x$.
    The minimum energy gap~$\Delta E$ increases monotonically with the driving strength (red curve), while the adiabaticity time~$T$, corresponding to the right-hand side of \cref{eq:adiabaticity}, exhibits a minimum at $h_x / J \approx 1.05$ (blue curve).
} 
\end{figure}

So far, we have chosen the strength $h^x$ of the driving Hamiltonian to be the same as the energy scale $J$ of the cost Hamiltonian, see \cref{subsubsec:energyscales}.
However, $h^x$ is actually a free parameter that can be tuned to optimize the annealing schedule.
In what follows, we discuss how the choice of the driving strength $h^x$ influences the size of the minimum energy gap and the annealing performance.

To this end, we introduce the dimensionless cost Hamiltonian $\Hcost^\prime = \Hcost / J$ and the dimensionless driving Hamiltonian $\Hdrive^\prime = \Hdrive / h^{x}$, expressed in units of the characteristic energy scale~$J$ and driving strength~$h^x$, respectively.
Consider now the dimensionless Hamiltonian $\hat{H}^\prime(g) = \Hcost^\prime + g \Hdrive^\prime$ and let us assume that the energy gap between the ground state and the first excited state of $\hat{H}^\prime(g)$ becomes minimal at the (\textit{a priori} unknown) critical value $g_c$.
Writing the instantaneous Hamiltonian for the linear annealing schedule in \cref{eq:annealingprotocol} as $\hat{H}(s) = s J \Hcost^\prime + (1 - s) h^x \Hdrive^\prime$, the relative strength $g_c$ between $\Hcost$ and $\Hdrive$, and thus the minimum energy gap, is reached at the time $s_c = (h^x / J) / (g_c + h^x / J)$ during the annealing sweep.
The instantaneous Hamiltonian at this time reads $\hat{H}(s_c) = s_c J \hat{H}^\prime(g_c)$.
Consequently, the minimum energy gap $\Delta E$ of the dimensionful Hamiltonian $\hat{H}$ relates to the dimensionless gap $\Delta E_c^\prime$ of $\hat{H}^\prime(g_c)$ as
\begin{align}
    \label{eq:gap_scaling_with_driving:J}
    \Delta E / J = s_c \Delta E_c^\prime = \frac{h^x / J}{g_c + h^x / J} \Delta E_c^\prime \,.
\end{align}
This relation shows that the minimum energy gap increases monotonically with the driving strength~$h_x$ and saturates at $J \Delta E_c^\prime$ for large $h_x / J$, as illustrated in \cref{fig:cost_to_driving_ratio} for a single 3-SAT instance.

Despite the monotonic behavior of the minimum energy gap, a larger driving strength does not necessarily result in a better annealing performance.
This is because a large value of $h_x / J$ results in a steep ramp, making the schedule less adiabatic.
Formally, this can be seen by examining the adiabatic condition in \cref{eq:adiabaticity}.
As shown in \cref{fig:cost_to_driving_ratio} for a single 3-SAT instance, the adiabaticity time reaches a minimum where $h_x$ and $J$ become of comparable strength.

To understand this behavior, it is instructive to write the adiabatic condition in \cref{eq:adiabaticity} as
\begin{equation}
    \label{eq:adiabaticitytradeoff}
    \sweepduration \gg T = \hbar \frac{V}{\Delta E^2} \approx \hbar \frac{J}{\Delta E^2} \Big( V_{\mathrm{cost}}^\prime + \frac{h^x}{J} V_{\mathrm{drive}}^\prime \Big) \,.
\end{equation}
Here, we have used the triangle inequality to separate the contributions of the cost and driving Hamiltonian to the quantity $V$ into $V_{\mathrm{cost}}^\prime = \max_s |\bra{1(s)} \Hcost^\prime \ket{0(s)}|$ and $V_{\mathrm{drive}}^\prime = \max_s |\bra{1(s)} \Hdrive^\prime \ket{0(s)}|$, assuming a linear annealing schedule.
This approximation becomes exact in the limit of either small or large $h^x / J$.
For small $h_x / J$, the contribution of $V_{\mathrm{drive}}^\prime$ can be neglected and the adiabaticity time is dominated by the behavior of the minimum energy gap according to \cref{eq:gap_scaling_with_driving:J}, causing the adiabaticity time to diverge for $h_x / J \to 0$.
For large $h_x / J$, the adiabaticity time is dominated by the linearly increasing second term in \cref{eq:adiabaticitytradeoff}.
The optimal driving strength corresponds to a tradeoff between these two scenarios, as illustrated in \cref{fig:cost_to_driving_ratio}.

As this discussion shows, for a linear annealing schedule, the driving strength $h^x$ should be chosen of comparable magnitude as the characteristic energy scale $J$ of the cost Hamiltonian.
The optimal value for $h_x$ depends on (generally unknown) problem characteristics. 
For the specific instance used in \cref{fig:cost_to_driving_ratio}, we find an optimal value of around $h_x/J\approx 1.05$. 
We checked that also for other problem instances the generic choice $h_x = J$ used throughout this work is not far from the optimal condition for adiabaticity. 

\section{Implementing PUBO via multi-qubit interactions in quantum hardware}
\label{sec:3body}

\begin{figure}[ht]
    \raggedright
    (a)\\ \includegraphics[width=0.6\columnwidth]{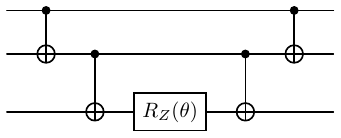}\\
    (b)\\ \includegraphics[width=\columnwidth]{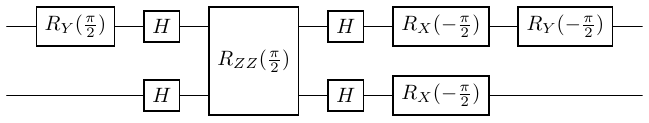}
    \\
    \subfloat{\label{fig:3-body_from_2-body:a}}%
    \subfloat{\label{fig:3-body_from_2-body:b}}%
    \caption{\label{fig:3-body_from_2-body}%
    (a) The three-qubit gate $R_{ZZZ}(\theta) = \exp(-i 4\theta \hat{s}_1^z \hat{s}_2^z \hat{s}_3^z)$ can be decomposed into four CNOT gates plus a single-qubit rotation. (b) Each CNOT gate can be further decomposed into  one $R_{ZZ}$ gate of fixed angle and multiple single-qubit rotations. The quantum gates in the decompositions are defined as $R_\alpha (\theta) = \exp(-i\theta \hat{s}^\alpha)$, with $\alpha \in \{X,Y,Z\}$, $R_{ZZ} (\theta) = \exp(- 2i\theta \hat{s}^z \hat{s}^z)$, and $H$ denoting the Hadamard gate.}
\end{figure}

In this section, we discuss possibilities to implement three-body interactions in order to realize third-order PUBO formulations in quantum hardware. The exploration of such multi-body interaction terms and their role in enhancing quantum annealing protocols, for example by using them as non-stoquastic driving terms~\cite{Ghosh2024}, has become an active area of research.
As we will see, the constant overhead of directly implementing PUBO problems is more than compensated---already for small problem sizes---by the advantageous scaling of the minimal gap investigated in the previous section.

\subsection{Native multi-qubit interactions and synthesis of three-body gates in quantum circuits\label{sec:3body:realization}}

Several proposals exist to directly engineer native three-body spin interactions of the type $\hat{s}_i^z \hat{s}_j^z \hat{s}_k^z$ on the hardware level. 
For example, in trapped-ion systems, these include concurrent off-resonant driving of first and second sideband~\cite{Bermudez2009} (also relevant for other three-body interactions~\cite{Yang2016,Andrade2022, Nagies2024b}) or a pulse sequence using squeezings and displacements of a phonon bus~\cite{Katz2022}. 
Another approach is via Floquet engineering of two-body interactions to obtain the three-body terms in a high-frequency expansion~\cite{Decker2020}.
In addition, proposals for effective multi-body interactions on superconducting qubit architectures have been put forward~\cite{Chancellor2017}.
Most available schemes share the common drawbacks that their implementation is somewhat complex and that the achievable higher-order interaction rates are considerably slower compared to those of two-body interactions.
Moreover, the tunability of the relative strengths of three-body interactions between different triples of spins depends on the utilized scheme as well as hardware characteristics.
Nonetheless, these techniques may be exploited in next-generation devices to realize analog quantum annealing with higher-order interactions that benefits from the advantages of direct PUBO encodings in terms of spatial and temporal resources.

While native higher-order interactions can also act as physical implementations of multi-qubit gates in digital quantum computers, due to the disadvantages mentioned above, most available platforms implement only single- and two-qubit gates natively.
These can realize higher-order interaction gates, required for PUBO formulations, with only polynomial overhead. 
An example is given in Fig.~\ref{fig:3-body_from_2-body}: 
a three-qubit gate of the form $R_{ZZZ}(\theta) = \exp(-i 4\theta \hat{s}_1^z \hat{s}_2^z \hat{s}_3^z)$ can be synthesized from four CNOT gates, plus a single-qubit rotation around the $z$-axis [see \cref{fig:3-body_from_2-body:a}].
Each CNOT gate can be further decomposed into
\begin{align}
\mathrm{CNOT}_{1,2} &=\mathrm{e}^{-i \pi/4}\nonumber\\
&\times R_{y_1}\left(-\frac{\pi}{2}\right) R_{x_1}\left(-\frac{\pi}{2}\right)R_{x_2}\left(-\frac{\pi}{2}\right)
 H_{1} H_{2} \nonumber\\
 &\times  R_{zz}\left(\frac{\pi}{2}\right) H_{1} H_{2}   R_{y_1}\left(\frac{\pi}{2}\right) \,,
 \end{align}
where we used a two-qubit controlled rotation gate $R_{zz}$, Hadamard gates $H$ and single-qubit rotations around $x$ and $y$-axis. We show the decomposition in \cref{fig:3-body_from_2-body:b}. The overall circuit for the three-qubit gate then uses four two-body gates of type $R_{zz}$ with fixed angle $\pi/2$ and 29 single-qubit rotations.
In this decomposition, two adjacent single-qubit gates could be further combined into only one single-qubit rotation.  
Similar decompositions are possible with other two-qubit gates like the Mølmer--Sørensen (MS) type gate $R_{xx}$.
Note that in digital quantum annealing protocols, the three-body interaction strength in each Trotter step is controlled by the rotation angle~$\theta$. Since $\theta$ enters in \cref{fig:3-body_from_2-body:a} only via a single $R_Z(\theta)$ gate, the overall three-qubit gate time is practically the same for different rotation angles.

As these examples show, the special structure of the $ZZZ$ gate enables implementations with far less resources than necessary for a general three-body unitary operation (for example, state of the art numerical sequential optimization requires 15 CNOT gates to decompose a general three-body unitary operation~\cite{Rakyta2021, Madden2022}). 
Importantly, while such an overhead 
may still be non-negligible in current noisy intermediate-scale quantum (NISQ) devices, it is a rather small constant factor.
As discussed below, this overhead can be more than compensated by the higher resource efficiency of the PUBO encoding.

\subsection{PUBO speedup}

As suggested by our numerical results in \cref{sec:numerics}, the minimal gap decreases more slowly with increasing system size for the PUBO formulation of certain 3-SAT problems than for the corresponding QUBO reduction, indicating that there exists a system size above which the PUBO formulation will always be favorable.
Specifically, this is the case for 3-SAT instances generated by \algorithm{toughSAT}, see \cref{fig:gapscaling:b}.

Even in scenarios where PUBO and QUBO scale approximately the same, as is the case for the 3-SAT problems generated by \algorithm{uniquePT1} shown in \cref{fig:gapscaling:a}, a constant PUBO speedup is still possible.
In the context of digital quantum annealing (see \cref{sec:annealing:digital}), we can use the estimated overhead of the three-body interactions from \cref{sec:3body:realization} together with the minimal gaps extracted numerically in \cref{sec:numerics} in order to estimate the potential gain in computing time when using the direct PUBO formulation according to \cref{eq:pubospeedup}.
In the example of \cref{fig:3-body_from_2-body}, executing the three-qubit gate involves four two-qubit gates and therefore takes about four times as long as a single two-qubit gate (neglecting single-qubit gate times).
In the context of quantum annealing, this corresponds to a ratio of three-body to two-body interaction strength of $J^{(3)} / J^{(2)}\approx 1/4$.
Using \cref{eq:pubospeedup} and the average values for the 
\algorithm{uniquePT1} generator listed in \cref{tab:minimalgap_fit}, we obtain
$\annealingtime_Q/\annealingtime_P \approx \num{12.8} \mathinner{(\tilde{V}_Q / \tilde{V}_P)} \etothe{\num{0.024} \, N}$.
In our numerics for small systems, we find a typical ratio of $\tilde{V}_Q / \tilde{V}_P \approx 0.5$, which only weakly depends on the system size. Our estimate therefore hints at a constant PUBO speedup in this regime; more precisely, an adiabatic sweep can be about six times faster if the PUBO instead of the QUBO formulation for this class of 3-SAT problems is used. 
If there is on top of that an (even small) exponential enhancement, as suggested by our numerical results, it will then dominate over a possible polynomial scaling of $\tilde{V}_Q / \tilde{V}_P$ at larger system sizes and further improve the performance of PUBO over QUBO.
Note that the precise value of the estimated speedup depends on the choice of the strength of the driving Hamiltonian as well as the strength of the constraint terms in the QUBO reduction. Using a better heuristic for the latter, e.g., the one proposed in Ref.~\cite{Alessandroni2023}, could possibly reduce the calculated PUBO improvement. From our analyses in \cref{subsubsec:penalty,subsec:drivingvscost}, where we found the used parameters to be close to optimal, we expect such a change to be only minor. 

For instances generated by \algorithm{toughSAT}, we see numerical evidence for a much stronger exponential scaling advantage of the PUBO formulation compared to the corresponding QUBO reduction: under the same assumptions as above, we find $\annealingtime_Q/\annealingtime_P \approx \num{0.81} \mathinner{(\tilde{V}_Q / \tilde{V}_P)} \etothe{\num{0.49} \, N}$. Already for small systems with $N=11$ classical variables, 
the adiabaticity time reduces by about two orders of magnitude in the PUBO formulation.
That is, solving this family of exponentially hard 3-SAT problems via quantum annealing is expected to be two orders of magnitude faster when using the PUBO formulation over the conventional approach of applying a QUBO reduction, and this estimate increases by another order of magnitude for every $\approx 4.7$ additional classical variables in the problem.

\section{Conclusion}
\label{sec:conclusion}

In this work, we have presented a variety of combinatorial problems that are naturally phrased in terms of higher-order polynomial unconstrained binary optimization. 
Using the PUBO formulation directly instead of the corresponding QUBO reduction---currently the standard choice in quantum annealing---can have a number of advantages: 
(1) PUBO formulations avoid additional ancillary qubits, thus reducing the spatial resources required. 
(2) The ancillary qubits required by QUBO reductions are typically subject to consistency constraints, which require large energy scales---a central limiting factor in existing quantum annealing hardware. 
(3) As we have illustrated numerically at the example of the paradigmatic 3-SAT problem, PUBO formulations can have significantly larger minimum energy gaps, allowing for faster annealing sweeps on both analog and digital platforms. 
This gain can outweigh the overhead incurred by higher-order interactions, as we have exemplified for digital quantum annealing with three-qubit gates synthesized from universal sets of single- and two-qubit gates.
In particular, we find an exponential advantage of PUBO over QUBO in the scaling of the minimum energy gap with the system size for certain classes of problems (in our example, already for small \algorithm{toughSAT} instances with $N=11$ variables, this suggests a potential speedup of about two orders of magnitude).
Altogether, implementing optimization problems in quantum annealers as native PUBO rather than QUBO can thus lead to savings in terms of both spatial and temporal resources.

On top of that, formulating certain problems as PUBO can reveal underlying structures and give insights into the complexity of certain problems.
As a case in point, rewriting 3-SAT instances generated by \algorithm{uniquePT4} (a generator constructing benchmark problems with unique solutions~\cite{Hsieh2021}) in PUBO form leads to a cost Hamiltonian described by only local fields that can be trivially solved.
In addition, our numerical analysis of the minimum energy gap reveals characteristics of other 3-SAT generators found in Ref.~\cite{Hsieh2021} (e.g., the fact that the generator \algorithm{uniquePT1} creates on average harder instances than \algorithm{toughSAT}), which are relevant for future studies of both classical and quantum algorithms where these generators can be used for benchmark purposes.

Although in this work we have focused on the paradigmatic 3-SAT problem, it is plausible that a similar exponential scaling advantage holds for other problems which are naturally described as a PUBO, such as those exemplified in \cref{app:other_PUBO}. 
In the future, it will be interesting to quantitatively study potential speedups also for these and other problems. 
Another interesting question is whether the presence of multi-qubit operators modifies the behavior of quantum resources during the annealing sweep~\cite{Orus2004,Lanting2014,Hauke2015,Santra2024}.
Finally, the favorable scaling of the adiabaticity time for certain families of 3-SAT problems, as suggested by our analysis of the minimum energy gap in \cref{subsec:scaling}, encourages more systematic studies comparing the time complexity of quantum annealing to state-of-the-art classical algorithms~\cite{Punnen2022,Glover2022,Du2025}.

\begin{acknowledgments}
We acknowledge useful discussions with Marcel Seelbach Benkner, Ivan Boldin, Michael Möller, Junichi Okamoto, Sebastian Rubbert,  Theerapot Sriarunothai and Christof Wunderlich.
The work reported in this publication is based on a project that was funded by the German Federal Ministry for Education and Research under the funding reference number 13N16437.
The authors are solely responsible for the content of this publication.
This work has benefited from Q@TN, the joint lab between University of Trento, FBK---Fondazione Bruno
Kessler, INFN---National Institute for Nuclear Physics, and CNR---National Research Council.
We acknowledge support by Provincia Autonoma di Trento.

This Accepted Manuscript is available for reuse under a CC BY-NC-ND licence after the 12 month embargo period provided that all the terms and conditions of the licence are adhered to.
\end{acknowledgments}

\appendix

\section{Selected PUBO problems} \label{app:other_PUBO}

There exists a plethora of optimization problems that are naturally posed as PUBO. In this appendix, we mention a few relevant examples.

\subsubsection{$p$-spin model}
The cost Hamiltonian of the $p$-spin model ($p\geq 2$) reads
\begin{equation}
\Hcost = -N \bigg( \frac{1}{N} \sum_{i=1}^N \hat{\sigma}_i^z \bigg)^p \,.
\end{equation}
For $p=2$, this Hamiltonian is known as the Lipkin--Meshkov--Glick model \cite{Lipkin1965, Meshkov1965, Glick1965}, while for $p\to\infty$ it recovers the Grover problem \cite{Joerg2010}. The solution of this model is trivial (all spin up for $p$ odd; all spin down or all spin up for $p$ even), and the large structure by which it is characterized enables efficient classical simulations even of non-stoquastic annealing protocols \cite{Ohzeki2017}. Nevertheless, its infinite range all-to-all interactions make it a useful toy model for testing improvements of quantum-optimization algorithms. 
For example, while standard annealing protocols with a transverse field as the driving Hamiltonian encounter a first-order phase transition in the $p$-spin model \cite{Joerg2010}, non-stoquastic protocols soften it to a second-order phase transition. This example illustrates the power of non-stoquastic protocols for improving adiabaticity and thus saving computing time \cite{Seki2012, Durkin2019}. 

\subsubsection{Hypergraph coloring}

Consider an undirected graph $G=(V,E)$, consisting of a finite set of vertices $V$ and a set of edges~$E$, the latter being a set of subsets of $V$. For a regular undirected graph, all edges are unordered pairs of vertices. For a hypergraph on the other hand, each edge can contain any number of vertices. 

Given $G$ and a number of $k$ colors, the hypergraph vertex coloring problem asks whether there is an assignment of exactly one colour to each vertex such that no edge contains vertices all of the same colour~\cite{Kang2023}. A QUBO formulation of the regular graph coloring is well-known~\cite{Lucas2014}, which can be easily generalized to hypergraphs: let $x^i_n = 1$ if vertex $n$ has color $i$, else $0$. The cost function of the problem can then be written as
\begin{align}
\label{eq:hypergraph}
    f(\vec{x}) &= \sum_{n \in V} \bigg(1 - \sum_{i=1}^k x_n^i \bigg)^2
    + \sum_{e \in E} \sum_{i=1}^k \prod_{n \in e} x^i_{n} \,,
\end{align}
where the first term ensures that each vertex has exactly one color assigned to it and the second term penalizes each edge whose vertices are all of the same color. Thus, any solution with zero energy is a proper vertex coloring of the hypergraph.

To implement this PUBO problem encoding on a quantum computer, $k |V|$ qubits are required (with $|V|$ being the number of vertices in the graph). To reduce \cref{eq:hypergraph} to a QUBO problem, one needs to introduce ancilla variables for every edge in the graph containing more than two vertices (the exact amount needed will depend on the degree of the edges and their overlap with each other).

Other generalizations of graph problems to hypergraphs similarly yield PUBO problems with higher-order interaction terms, e.g., the vertex cover problem or the graph partitioning problem~\cite{Dominguez2023}.

\subsubsection{NAE-k-SAT}

Apart from $k$-SAT, which we have discussed in detail in \cref{subsec:3sat}, there are many other variations and constraints on the general satisfiability problem. One of those is the NAE-$k$-SAT (Not-All-Equal-k-SAT) problem~\cite{Bashar2023}, which can also be naturally formulated as a PUBO. Like $k$-SAT, NAE-$k$-SAT consists of a conjunction of clauses, where each clause contains $k$ Boolean variables and is only \textit{true}, if at least one variable is \textit{true} \emph{and} at least one variable is \textit{false}. Only for $k \geq 4$ is the problem naturally formulated as a PUBO with higher-order interactions~\cite{Bashar2023}. For example, for $k=4$, a clause $C(x_i,x_j,x_k,x_l)$ can be encoded into one four-body interaction $\sim \hat{s}^z_i \hat{s}^z_j \hat{s}^z_k \hat{s}^z_l$ and six two-body interaction penalty terms of the cost Hamiltonian.

As for $k$-SAT, this PUBO encoding requires one qubit for each Boolean variable. The equivalent reduction to a QUBO, on the other hand, requires (for $k=4$) in the worst case two additional ancilla qubits for each clause in the problem. 

\subsubsection{Traveling salesperson problem with time windows}

A variation on the famous traveling salesperson problem (TSP) is the traveling salesperson problem with time windows (TSPTW). Here, like in the TSP, we ask for a Hamiltonian path with minimum weight on a graph with weights assigned to all edges. In addition, each vertex has to be visited in a specific time window. Both QUBO and PUBO formulations of this problem are known. In Ref.~\cite{Salehi2022}, the authors propose two different QUBO and one PUBO formulation of degree four for the problem.
In the PUBO case, the number of required qubits scales as $\mathcal{O}(n^2 + n\delta)$ in the number of vertices $n$ (here, $\delta$ is a number scaling at most linearly in $n$). In contrast, the QUBO formulations scale as $\mathcal{O}(n^3 + n\delta)$ and $\mathcal{O}(n^2 + n^2\delta)$, respectively, for the edge-based and integer linear programming (ILP) formulations.

\section{QUBO formulations of 3-SAT} \label{app:QUBO_reducs}

The most resource-efficient way (in terms of qubit requirements) to formulate a 3-SAT instance as a QUBO problem is the direct reduction of the PUBO formulation to QUBO via the introduction of slack variables (see \cref{subsubsec:3satquboreduc}). Here, we also list two more resource-intensive methods.

\subsubsection{3-SAT as QUBO via the maximum independent set problem}

A well-known example of a QUBO formulation of 3-SAT uses the equivalent maximal independent set problem (MIS)~\cite{Choi2010}: Draw a graph with three vertices for each clause (corresponding to the three involved literals). All vertices belonging to the same clause as well as pairs of vertices corresponding to mutually exclusive literals (i.e., $x_{i}$ and $\lnot{x_i}$) are connected by edges, the set of which is denoted with $E$. The MIS problem then asks whether it is possible to color $M$ vertices without two colored vertices being connected by an edge. If the answer is yes, the original 3-SAT problem is satisfiable. This MIS problem can be easily translated to a QUBO problem requiring $3M$ qubits (or about $\round(3 \times 4.24 \, N) = \round(12.72 \, N)$ for hard problems):
\begin{align}
    f_{\mathrm{MIS}}(\vec{x}) = \sum\limits_{ij \in E}x_{i}x_j \,.
\end{align}
Although generically more resources are required than for the reduction with slack variables, a potential advantage of the MIS formulation, especially regarding its use on physical quantum devices, lies in the fact that all interactions have the same energy scale, whereas slack variables for a direct QUBO reduction are introduced with constraints that are imposed through high energy penalties.

\subsubsection{3-SAT as QUBO via linear inequality constraints}

Finally, we mention also a linear formulation of 3-SAT as a QUBO problem \cite{Herrman2021}, where one introduces one ancillary Boolean variable $z_{m}$ for each clause $C_m$ that indicates whether said clause is fulfilled or not.
One then asks to maximize the number of fulfilled clauses
\begin{align}
    \text{max}\sum_m z_m
\end{align}

under the inequality constraints 
\begin{align}
    \sum\limits_{x_{i} \in \mathrm{TRUE}_{m}}x_{i}+\sum\limits_{x_{i} \in \mathrm{FALSE}_{m}} (1-x_{i})\geq z_m \,,
\end{align}
where $\mathrm{TRUE}_m$ ($\mathrm{FALSE}_m$) is the set of literals contained in $C_m$ satisfying the clause if chosen as $\textit{true}$ ($\textit{false}$). 
One can then introduce two slack variables per inequality constraint to express them as equalities and write the problem in QUBO form. 

Overall, this approach needs $3M$ ancilla qubits in addition to the $N$ qubits for the literals, as well as an energy penalty to enfore the constraints. 
This scaling has to be compared to the one using the mapping to the maximal independent set problem ($3M$ qubits, no different energy scales), the one decomposing the PUBO problem using slack variables ($N$ qubits plus at most $M$ ancillas, plus energy penalties), and the direct PUBO implementation ($N$ qubits, interactions of degree three, no  different energy scales).

\bibliography{trappedions2}

\end{document}